\documentclass[twocolumn,aps]{revtex4}
\usepackage{amsmath}
\usepackage{amssymb}
\usepackage{amsfonts}
\usepackage{bm}
\usepackage[dvips]{graphicx}
\usepackage{dcolumn}
\usepackage{txfonts}
\usepackage{makeidx}
\usepackage{color}
\usepackage{mathtools}
\usepackage{threeparttable}
\usepackage[linkcolor=blue,anchorcolor=black,citecolor=blue]{hyperref}

\newcommand{\tmop}[1]{\ensuremath{\operatorname{#1}}}

\begin{document}

\title{Quantum network of superconducting qubits through opto-mechanical
interface}

\author{Zhang-qi Yin$^{1}$, W. L. Yang$^{2}$, L. Sun$^{1}$, L. M. Duan$^{3,1}$}

\affiliation{$^1$Center for Quantum Information, IIIS, Tsinghua University, Beijing
100084, P. R. China}

\affiliation{$^2$State Key Laboratory of Magnetic Resonance and Atomic and Molecular
Physics, Wuhan Institute of Physics and Mathematics, Chinese Academy
of Sciences, Wuhan 430071, China}

\affiliation{$^3$Department of Physics, University of Michigan, Ann Arbor, Michigan
48109, USA}

\begin{abstract}
We propose a scheme to realize quantum networking of superconducting
qubits based on the opto-mechanical interface. The superconducting
qubits interact with the microwave photons, which then couple to the
optical photons through the opto-mechanical interface. The interface
generates a quantum link between superconducting qubits and optical
flying qubits with tunable pulse shapes and carrier frequencies, enabling
transmission of quantum information to other superconducting or atomic
qubits. We show that the scheme works under realistic experimental
conditions and it also provides a way for fast initialization of the
superconducting qubits under $1$ K instead of $20$ mK operation
temperature.
\end{abstract}

\date{\today}

\maketitle
\section{Introduction}
Superconducting qubits (SQs) constitute one of the leading candidate
systems for realization of quantum computation \cite{Science2013}.
Through the circuit resonators, SQs have strong coupling to the microwave
photons \cite{Science2013}, which can be used
for qubit interaction, state engineering of the photonic modes, and
non-destructive readout of the qubits \cite{Wallraff04,You05}. Universal
quantum logic gates have been realized for SQs in circuit QED (cQED)
systems with high fidelity and speed \cite{Barends14}. Through use
of the noise insensitive qubits, the coherent time of the SQs has
been increased by several orders of magnitude in recent years and
pushed to the $100~\mu$s region \cite{Paik11,Chang13}. 
In a single circuit resonator, the number of SQs is still limited.
Further scaling up the number of qubits requires linking distant cQED
systems to form a quantum network. Microwave photons are sensitive
to thermal noise and their quantum states only survive under cryogenic
temperature. So it is hard to use them to link SQs in two different
setups. Optical photons, on the other hand, are robust information
carriers at room temperature and serve as ideal flying qubits for
long-distance communication. They can carry quantum information to
distant locations through an optical fiber.

In this paper, we propose a scheme to realize a quantum network of
SQs through an opto-mechanical interface that couples optical photons
in a cavity to microwave photons and SQs in a circuit resonator. The
interface generates entangled states between SQs and photonic pulses
with tunable pulse shape and carrier frequency. The photons then make
a quantum link between distant SQs through either a measurement-based
entangling protocol or a deterministic state mapping. Because of the
tunability of shape and frequency of the emitted photon, the same
scheme can also be used to realize a hybrid network between SQs and
other matter qubits such as atomic ions \cite{Olmschenk09}, quantum
dots \cite{Schaibley13}, or defect spins in solids \cite{Bernien13}.
A hybrid network may allow combination of advantages of different
kinds of qubits. For instance, SQs may be good for fast information
processing while atomic qubits are ideal for quantum memory. Our scheme
is based on the recent advance on the microwave-optical interface:
there have been several proposals to realize this interface with ions
\cite{Kielpinski12,Hafezi12}, cold atoms \cite{Verdu09}, or a hybrid opto-mechanical
system with superconducting resonators \cite{Regal11,Taylor11,Barzanjeh12,Wang12,Tian12,Xue13,Zhang14},
or with flux qubit \cite{Xia04}.
In particular, a recent experiment has demonstrated the transducer
between microwave and optical photons using the opto-mechanical system
at $4.5$K temperature \cite{Andrews13}. One hassle for an interface
between SQs and optical photons is that thermal initialization of
the SQs requires an operating temperature around $20$ mK in a dilution
fringe, while an interface to photons requires an optical window,
which introduces heating due to black-body radiation and may significantly
increase the system temperature. We circumvent this problem by showing
that our proposed scheme can achieve fast initialization of the SQs
at $1$ K through optical sideband cooling by use of the same opto-mechanical
interface.

\begin{figure}[htbp]
\centering \includegraphics[width=5.5cm]{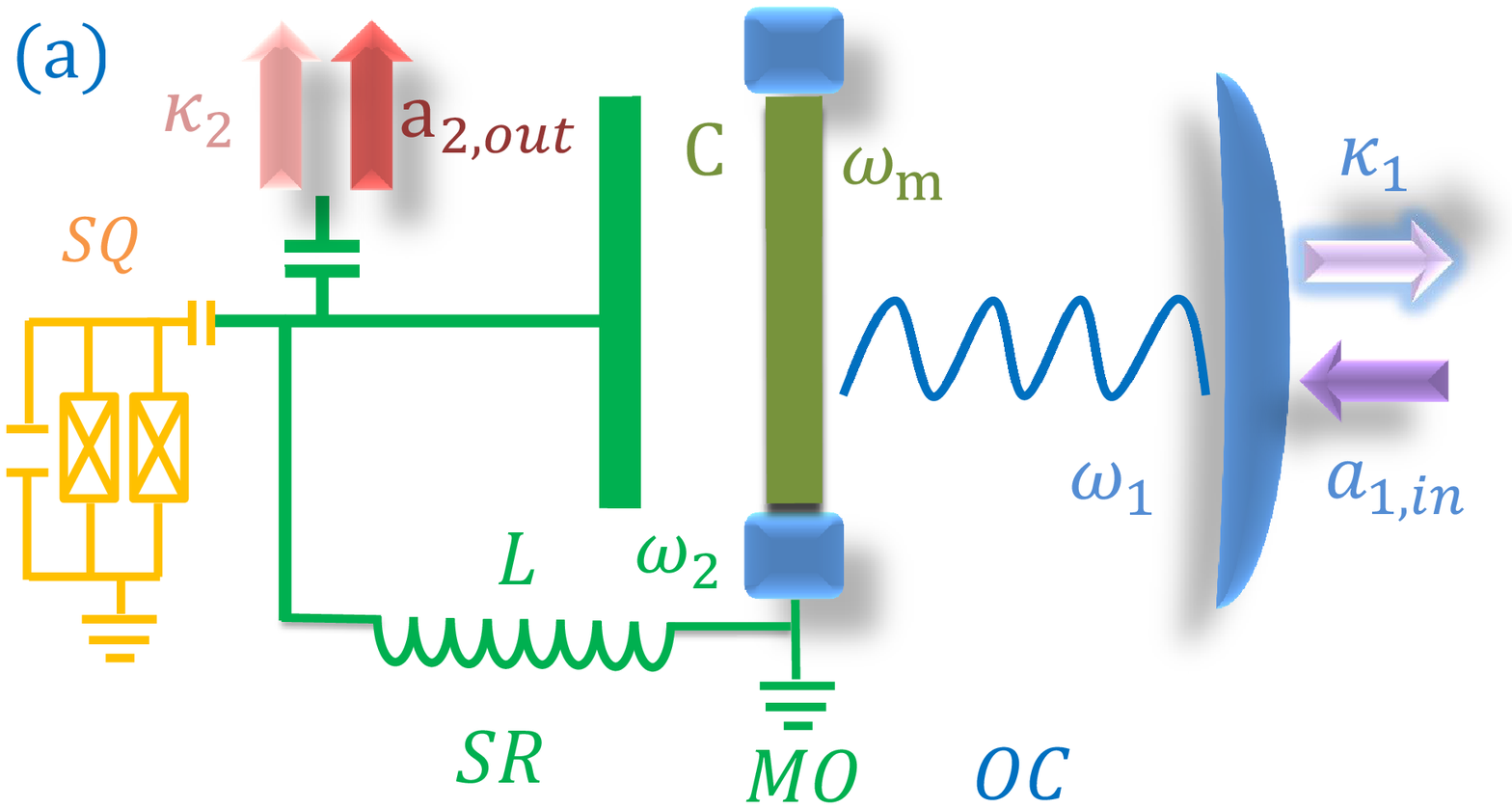} \includegraphics[width=1.9cm]{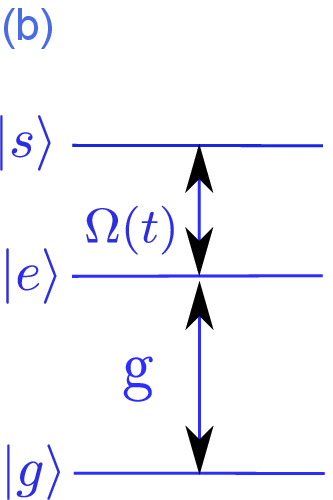}
\caption{(Color online) (a) The schematic scheme of the opto-mechanical quantum interface.
The SQ couples with the microwave mode
$a_{2}$ in a superconducting resonator (SR). The mechanical oscillator
(MO) mode $a_{m}$ for vibration of the interface couples simultaneously
to the mode $a_{2}$ of the SR and the mode $a_{1}$ of the optical
cavity (OC). Both $a_{2}$ and $a_{1}$ modes are driven by coherent
classical fields on the red sideband. (b) The energy levels of the
superconducting junction, where $|g\rangle$ is the ground state,
$|e\rangle$ is the first excited state, and $|s\rangle$ is the second
excited state. The transition $|g\rangle$ to $|e\rangle$ couples
to the mode $a_{2}$ with coupling rate $g_{c}$, while transition
$|e\rangle$ to $|s\rangle$ is driven by a microwave field with Rabi
frequency $\Omega\left(t\right)$.}
\end{figure}

\section{The Model}
As show in Fig. 1, the system we consider
contains an optical cavity (OC) and a microwave superconducting resonator
(SR) \cite{Ong11,Dewes12}, which share an interface that can vibrate and forms a mechanical
oscillator (MO) \cite{Teufel11a,Teufel11b}. The shared vibrating interface
between the OC and the SR has been proposed in several schemes \cite{Regal11,Taylor11,Barzanjeh12,Wang12,Tian12}
and realized very recently in experiments \cite{Andrews13,Suchoi14}. For this system,
the MO mode $a_{m}$ of frequency $\omega_{m}$ couples simultaneously
to the optical mode $a_{1}$ of frequency $\omega_{1}$ and the microwave
mode $a_{2}$ of frequency $\omega_{2}$. We have assumed that the
coupling rate is much less than the mode spacing of either of these
oscillators so that only one mode is relevant respectively for the
OC, the MO, and the SR. The optical and the microwave modes $a_{1}$
and $a_{2}$ are driven at the red-sideband with frequency $\omega_{L1}=\omega_{1}-\Delta_{1}$
and $\omega_{L2}=\omega_{2}-\Delta_{2}$, respectively. We set $\Delta_1=
\Delta_2=\omega_m$. Inside the
SR, there are nonlinear Josephson junctions, with the lowest three
anharmonic levels shown in Fig. 1b. The
levels $|g\rangle$ and $|s\rangle$ make a SQ, with coupling mediated
by the middle level $|e\rangle$ with a coupling rate $g_{c}$ for
the $|g\rangle\rightarrow|e\rangle$ transition and a Rabi frequency
$\Omega\left(t\right)$ (driven by a microwave field with tunable
shape) for the $|e\rangle\rightarrow|s\rangle$ transition.

The Hamiltonian of the system has the form $H=H_{0}+H_{I}+H_{d}$,
where $H_{0}=\sum_{i=1,2}\omega_{i}a_{i}^{\dagger}a_{i}+\omega_{m}a_{m}^{\dagger}a_{m}+\omega_{e}\sigma_{ee}$,
$H_{I}=\sum_{i=1,2}g_{i}a_{i}^{\dagger}a_{i}(a_{m}+a_{m}^{\dagger})+g_{c}(\sigma_{eg} +\sigma_{ge})(a_{2}+a_{2}^\dagger)$,
and $H_{d}=\sum_{i=1,2}(\frac{\Omega_{i}}{2}e^{-i\omega_{Li}t}+ \mathrm{h.c.})(a_i+a_i^\dagger)+ (\frac{\Omega'}{2}
e^{-i\omega_{L_2} t} + \mathrm{h.c.})(\sigma_{ge}+\sigma_{eg})$.
We have set $\hbar=1$ and taken the notation $\sigma_{\mu\nu}=|\mu\rangle\langle\nu|\;(\mu,\nu=g,e,s)$.
The SQ and SR drive pulses are generated by two phase-locked microwave generators.
The flux control pulses are used to tune the SQ to be resonant with the SR with $\omega_{2}=\omega_{e}$ \cite{Dewes12}.
The opto-mechanical coupling
rates $g_{i}$ ($i=1,2$) are typically small, but their effect can
be enhanced through the driving field $\Omega_{i}$. Under the driving,
the steady state amplitude of the mode $a_{i}$ is given by $\alpha_{i}\approx\Omega_{i}/2\Delta_{i}$.
We take the driving strength $\Omega'^*=g_c\Omega_2/\omega_m$.
The opto-mechanical coupling terms can be expanded with $a_{i}-\alpha_{i}$
and the effective coupling Hamiltonian takes the form
(see details in Appendix \ref{appendixA})\cite{Wang12,Tian12,Yin09b}
\begin{equation}
 \begin{aligned}
H_{om}=&\sum_{i=1,2}\left[\omega_m a_{i}^{\dagger}a_{i}+G_{i}(a_{i}^{\dagger}+a_{i})(a_{m}+
a_{m}^{\dagger})\right] \\&+\omega_{m}a_{m}^{\dagger}a_{m} +
( g_c a_2 \sigma_{eg} +\mathrm{h.c.})
 \end{aligned}
\end{equation}
where $G_{i}=\alpha_{i}g_{i}$.  Under the rotating wave approximation ($\omega_{m}\gg G_{i}, g_c$),
the whole Hamiltonian in the interaction picture is given by
\begin{equation}
H_{I}=\left(G_{1}a_{1}^{\dagger}+G_{2}a_{2}^{\dagger}\right)a_{m}+g_{c}\sigma_{eg}a_{2}+\mathrm{h.c.}.\label{eq:Heff}
\end{equation}
The corresponding Langevin equations for the $a_{j}$ ($j=1,2,m$)
modes and the SQ take the form
\begin{equation}
\begin{aligned}\dot{a}_{j}= & -i[a_{j},H_{I}]-\frac{\kappa_{j}}{2}+\sqrt{\kappa_{j}}a_{j}^{\mathrm{in}},\\
\dot{\sigma}_{ge}= & -i[\sigma_{ge},H_{I}],-\frac{\gamma}{2}\sigma_{ge}+\sqrt{\gamma}\sigma_{z}a_{s}^{\mathrm{in}},
\end{aligned}
\label{eq:Langevin}
\end{equation}
where $\sigma_{z}=\sigma_{ee}-\sigma_{gg}$, $\gamma$ is the decay
rate of the level $|e\rangle$, and $\kappa_{j}$ is the decay rate
of the mode $a_{j}$.

\section{SQ intialization and SQ-photon quantum interface}
 Without loss of generality, we take $G_{1}=G_{2}=G$ for simplicity of notation.
We may define the normal modes $b,b_{\pm}$ with $a_{1}=(b_{+}+b_{-}-\sqrt{2}b)/2$,
$a_{2}=(b_{+}+b_{-}+\sqrt{2}b)/2$, $a_{m}=(b_{+}-b_{-})/\sqrt{2}$,
which diagonalize the opto-mechanical coupling Hamiltonian \cite{Yin07}.
The SQ only resonantly couples with normal mode $b$.
The normal mode $b$ decays through two channels, $a_{1}^{\mathrm{out}}$
and $a_{2}^{\mathrm{out}}$. The decay of $b$ mode is denoted as
$\kappa=(\kappa_{1}+\kappa_{2})/2$. Typically, we have $\kappa_{1}\gg\kappa_{2}$,
so the photons go out dominantly through the $a_{1}^{\mathrm{out}}$
channel, which is vacuum. As the SQ only strongly couples with the normal mode $b$,
the steady state of SQ will approach to the ground state $|g\rangle$.
If the SQ is initially in a mixture
of $|g\rangle$ and $|e\rangle$ states, we can cool
it to the ground state $|g\rangle$ by driving the red sideband
of the optical cavity \cite{Wilson07,Marquardt07,Yin09a,Yin11,Liu13}.
 If the initial state of the SQ involves mixture of other states,
these other states can be first driven to the state $|e\rangle$ through a
microwave filed and then decay to the ground state $|g\rangle$ by the
opto-mechanical sideband cooling. The working
temperature temperature for both initialization and interface can be much higher than tens of mK.

In order to couple the
SQ to an output optical photon with controllable pulse shape, we prepare
the SQ initially on the level $|s\rangle$ and drive the transition
$|s\rangle$ to $|e\rangle$ by a microwave field with Rabi frequency
$\Omega(t)$ and pulse duration $T_D$. The total Hamiltonian of the
system is $H_{t}=H_{I}+(\Omega(t)\sigma_{se}+\mathrm{h.c.})$.
In the limit $T_{D}^{-1}\ll G,g,\kappa_{1}$, the modes $b_{\pm}$
are not populated and can be adiabatically eliminated. The effective
Hamiltonian is simplified to $H_{t}=\Omega(t)\sigma_{se}+\frac{\sqrt{2}}{2}g_{c}b\sigma_{eg}+\mathrm{h.c.}$
The Hamiltonian $H_{t}$ has a dark state $|D\rangle=[|s\rangle|0\rangle-r(t)|g\rangle|1\rangle]/\sqrt{1+|r(t)|^{2}}$,
where $r(t)=\sqrt{2}\Omega(t)/g_{c}$, and $|0\rangle,|1\rangle$
represent the Fock states of the mode $b$.
To solve the output pulse shape, we rewrite the dark state
as $|D\rangle=\cos\theta|s\rangle|0\rangle-\sin\theta|g\rangle|1\rangle$,
with $\cos\theta=1/\sqrt{1+|r|^{2}}$, and define an orthogonal bright
state $|B\rangle=\sin\theta|s\rangle|0\rangle+\cos\theta|g\rangle|1\rangle$.
The wave-function of the whole system can be expanded as $|\Psi\rangle=(c_{d}|D\rangle+c_{b}|B\rangle+c_{e}|e\rangle)\otimes|\mathrm{vac}\rangle+|g\rangle|0\rangle\otimes|\varphi\rangle$,
where $|\mathrm{vac}\rangle$ is the vacuum state of output field,
and $|\varphi\rangle=\int_{-\omega_{c}}^{+\omega_{c}}d\omega c_{\omega}a_{out}^{\dagger}(\omega)|\mathrm{vac}\rangle$
denotes the single-photon state of the output filed with frequency
spectrum $c_{\omega}$. The dynamics of system is determined by the
Schrödinger equation $i\partial_{t}|\Psi\rangle=H_{t}|\Psi\rangle$,
where $H_{t}$ is the total Hamiltonian that includes the input-output
coupling terms \cite{Duan03}. Using the method in Ref. \cite{Duan03},
the output pulse shape $f(t),$ given by the Fourier transform of
$c_{\omega}$, can be solved analytically in the adiabatic limit,
with

\begin{equation}
f(t)=\sqrt{\kappa}\sin\theta\exp(-\frac{\kappa}{2}\int_{0}^{t}\sin^{2}\theta(\tau)d\tau).\label{eq:output}
\end{equation}
The pulse shape $f(t)$ is fully determined by $\theta(t)$.

\begin{figure}[htbp]
\centering 
 \includegraphics[width=4.1cm]{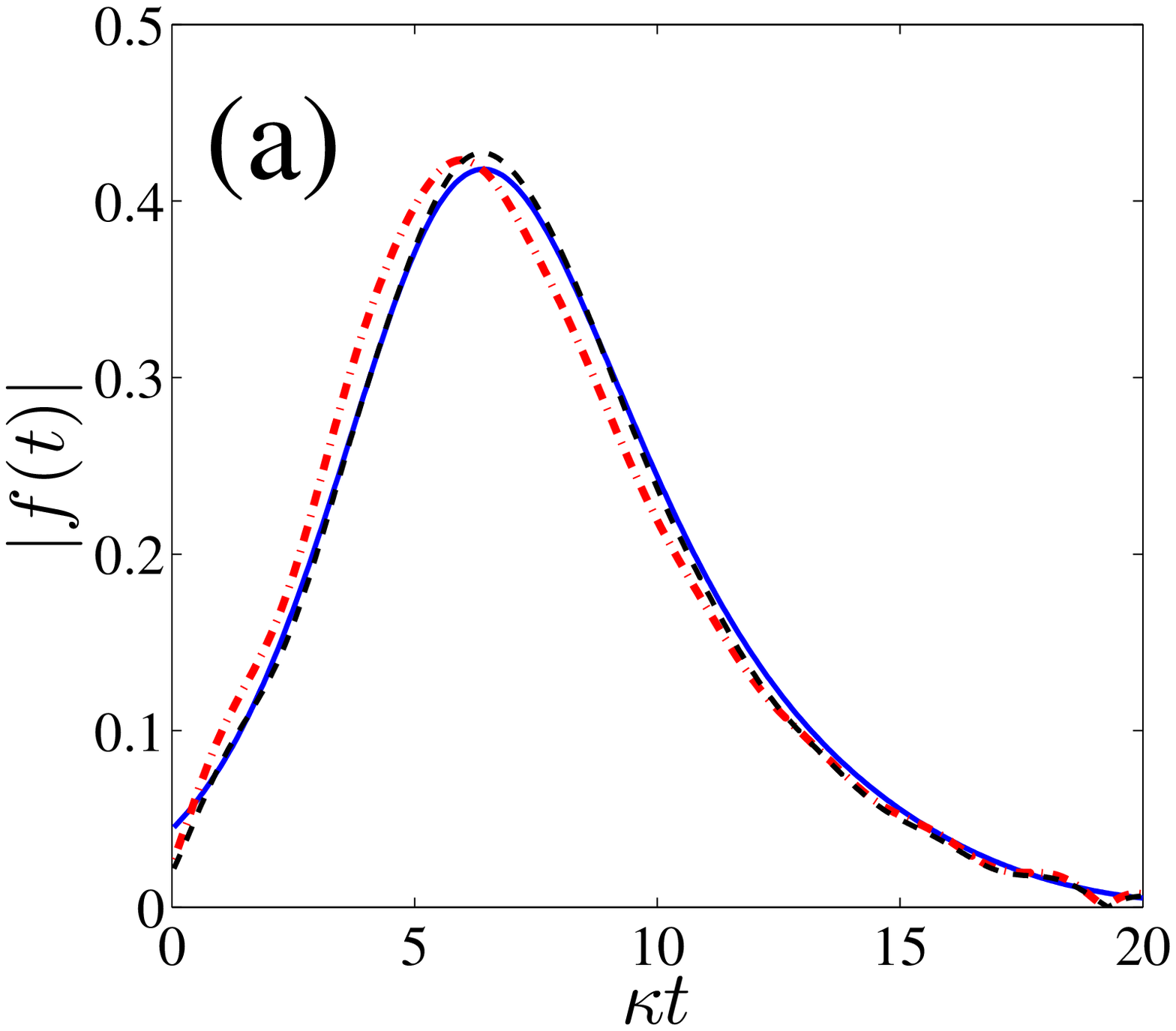} \includegraphics[width=4cm]{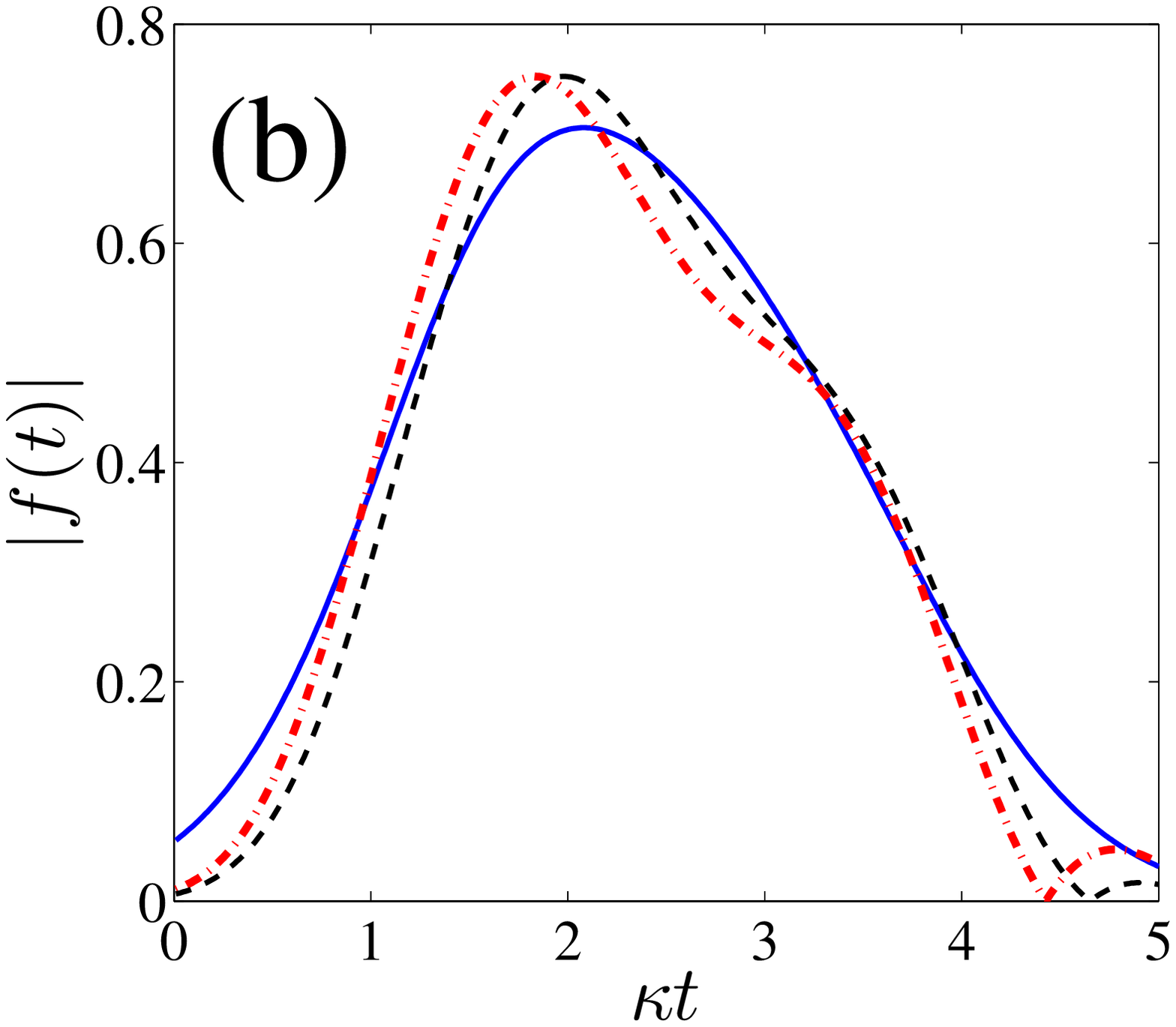}
\includegraphics[width=4.1cm]{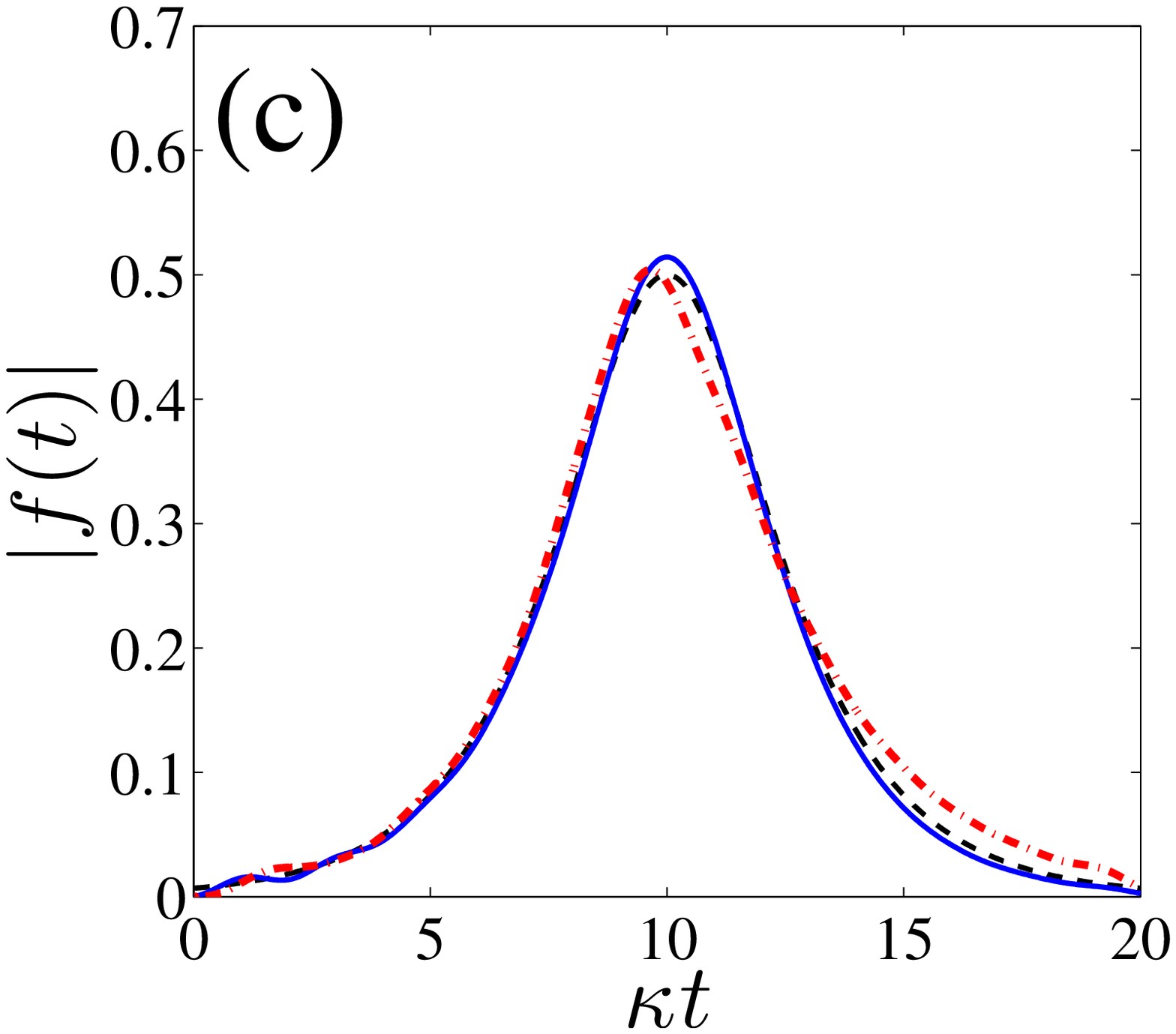} \includegraphics[width=4cm]{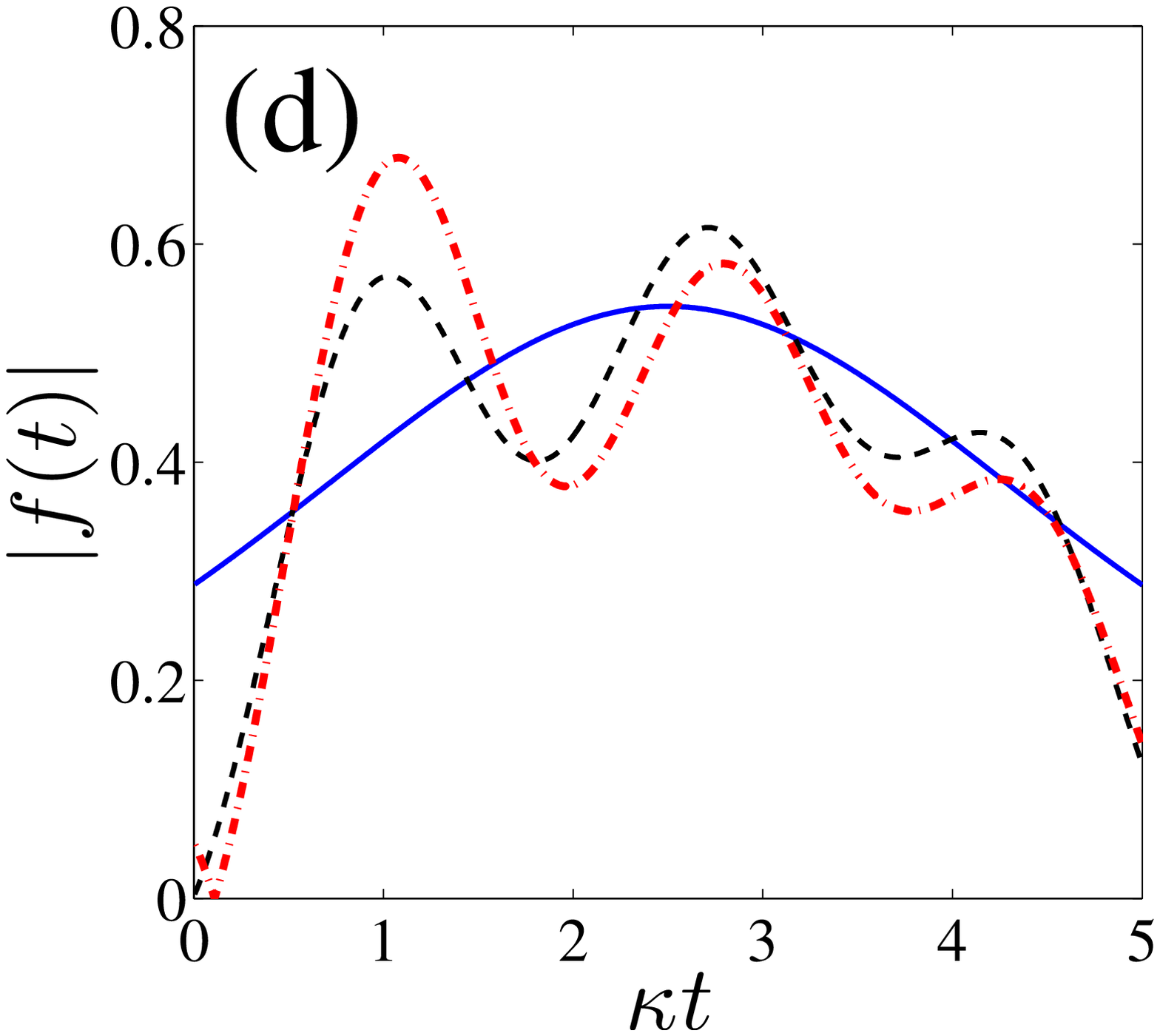}
\caption{(Color online) (a) The shape of the output single-photon pulse $|f(t)|$. We take
$g=G=3\kappa$ and the pulse duration $T_{D}=20/\kappa$. The driving
pules $\Omega(t)=ge^{-(t-T_D/2)^2/2t_w^2}$ is assumed to be a Gaussian shape with the peak
at $t=T_{D}/2$ and a width $t_{w}=T_{D}/5$. The solid (dash, dash-dot)
curve represents respectively the analytic pulse shape in Eq. (4)
derived in the adiabatic limit (the numerical result that includes
contribution of the bright state $|B\rangle$, the exact result that
includes contributions of all the modes $b,b_{\pm}$). The shape function
is normalized according to $\int|f(t)|^{2}dt=1$ for the convenience
of comparison. The overlap between the exact shape (dash-dot curve) and
the adiabatic shape (solid curve) is about $99\%$. (b) Same as Fig.
(a) but with the pulse duration $T_{D}=5/\kappa$. The adiabatic approximation
is not well satisfied in this case, and the shape overlap is reduced
to $80\%$. (c) Same as Fig. (a) but wih the driving Rabi frequency
$\Omega(t)=\left(g_{c}/\sqrt{2}\right)e^{\kappa(t-T_D/2)/2}$, which
gives a symmetric output pulse shape \cite{Duan03}. In the adiabatic
limit, the shape (the solid curve) is given by the analytic form $f(t)=\sqrt{\kappa/4}sech\left[\kappa(t-T_D/2)/2\right]$,
which has overlap of $99.7\%$ with the exact shape. (d) Same as Fig.
(c) but with the puse duration $T_{D}=5/\kappa$. }

\label{fig:output}
\end{figure}

To check whether the pulse shape of Eq. (4) derived under the adiabatic
limit holds under typical experimental parameters, we compare in Fig.
(2) the pulse shapes obtained from the analytic formula and from the
exact numerical simulation. In numerical simulation, we solve the
exact system dynamics by including the contribution of populations
either in the bright state $|B\rangle$ or of all the three modes
$b$ and $b_{\pm}$. As one can see from Fig. \ref{fig:output}, if
the pulse duration $T_{D}\gtrsim20/\kappa$, the pulse shape from
the analytic formula (4) overlaps very well with the exact result,
with the mismatching error less than $1\%$. However, for a short
pulse with $T_{D}\sim5/\kappa$, there is a significant shape mismatching
error and one should use the exact result instead of the approximate
analytic formula. The exact result shows some oscillations in the
pulse shape for a short driving field, resulting from the population
oscillation in different modes $b$,$b_{\pm}$ when the condition
of adiabatic elimination $T_{D}^{-1}\ll G,g,\kappa_{1}$ is not well
satisfied.

\section{Quantum networking of SQs}
 In the above, we have shown
how to couple a SQ to a single optical output photon with a controllable
pulse shape. This ability is critical for building up a quantum network
of SQs or a hybrid network between SQs and other matter qubits. Here,
we mention two complementary schemes for quantum networking of SQs,
requiring different kinds of pulse shape control.

The key requirement of quantum networking is to generate entanglement
between remote SQs. The first scheme for entanglement generation is
based on a deterministic quantum state transfer between SQs in two
remote cavities \cite{Cirac97}. As absorption is the time reversal
of the emission process, it has been shown in Ref. \cite{Cirac97}
that an emitted single-photon pulse can be completely absorbed by
a matter qubit in a cavity if we simultaneously reverse the temporal
shape of the photon pulse and the driving filed $\Omega(t)$. As shown
in Fig. 2, with an appropriate control of the driving microwave field
$\Omega(t)$, we can transfer a quantum state from a SQ to a single-photon
pulse with a symmetric temporal shape. This single-photon pulse, after
propagation in an optical fiber, can then be absorbed by a SQ in another
remote cavity, if the driving $\Omega'(t)$ of the second SQ is the
time reversal of $\Omega(t)$. The shape control of the driving microwave
pulse $\Omega(t)$ or $\Omega'(t)$ can be easily achieved through
modulation by an arbitrary wave form generator. If we make a half
transfer of the population from the first SQ to the photonic pulse,
the generated state between the SQ and the output photon $p$ has
the form $\left(|s\rangle_{1}|0\rangle_{p}+|g\rangle_{1}|1\rangle_{p}\right)/\sqrt{2}$.
Then, after absorption of the photon by the second SQ, we generate
an entangled state $\left(|s\rangle_{1}|g\rangle_{2}+|g\rangle_{1}|s\rangle_{2}\right)/\sqrt{2}$
between two remote SQs, as required for quantum networking.

\begin{figure}[htbp]
\centering \includegraphics[width=4.4cm]{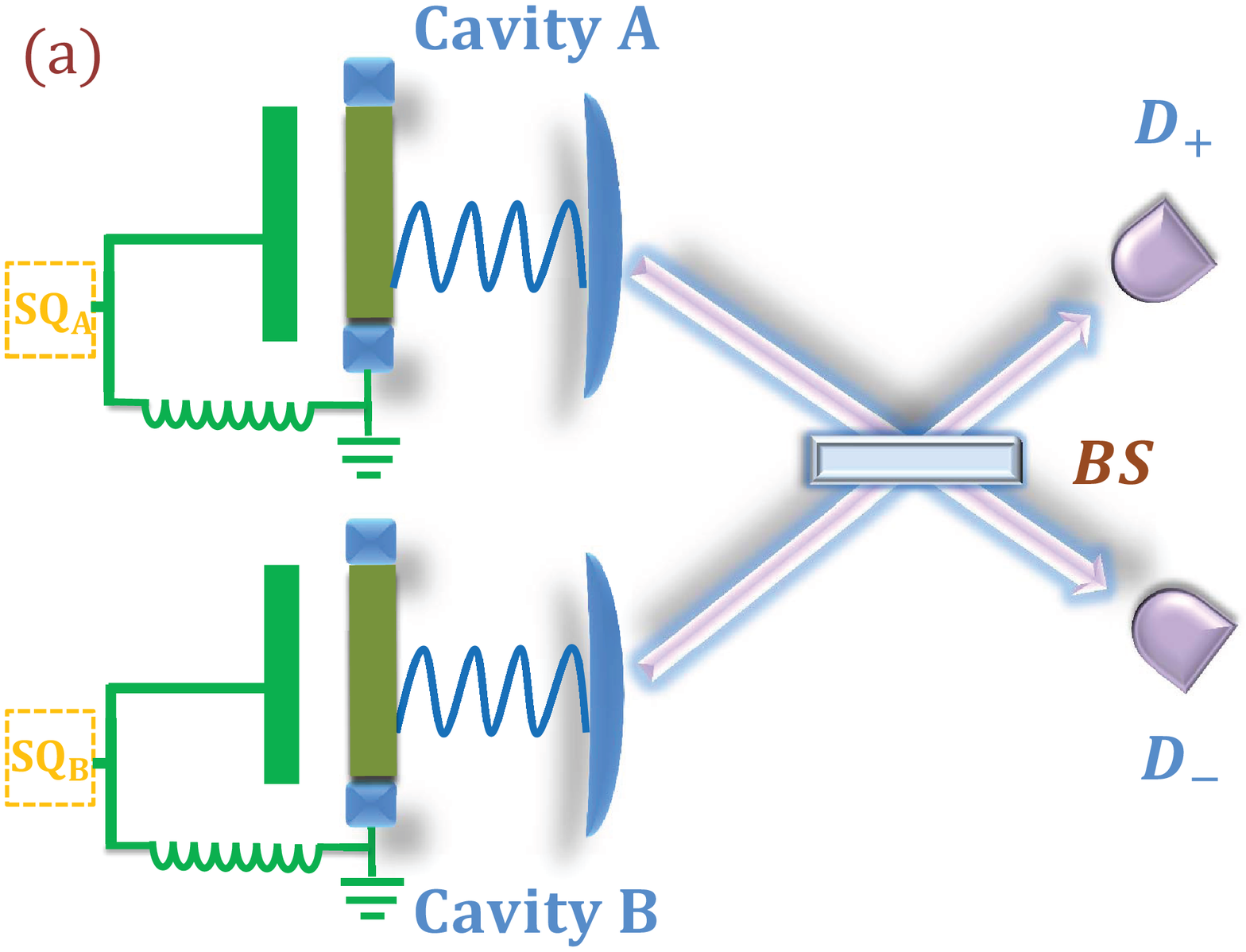}
\includegraphics[width=3.5cm]{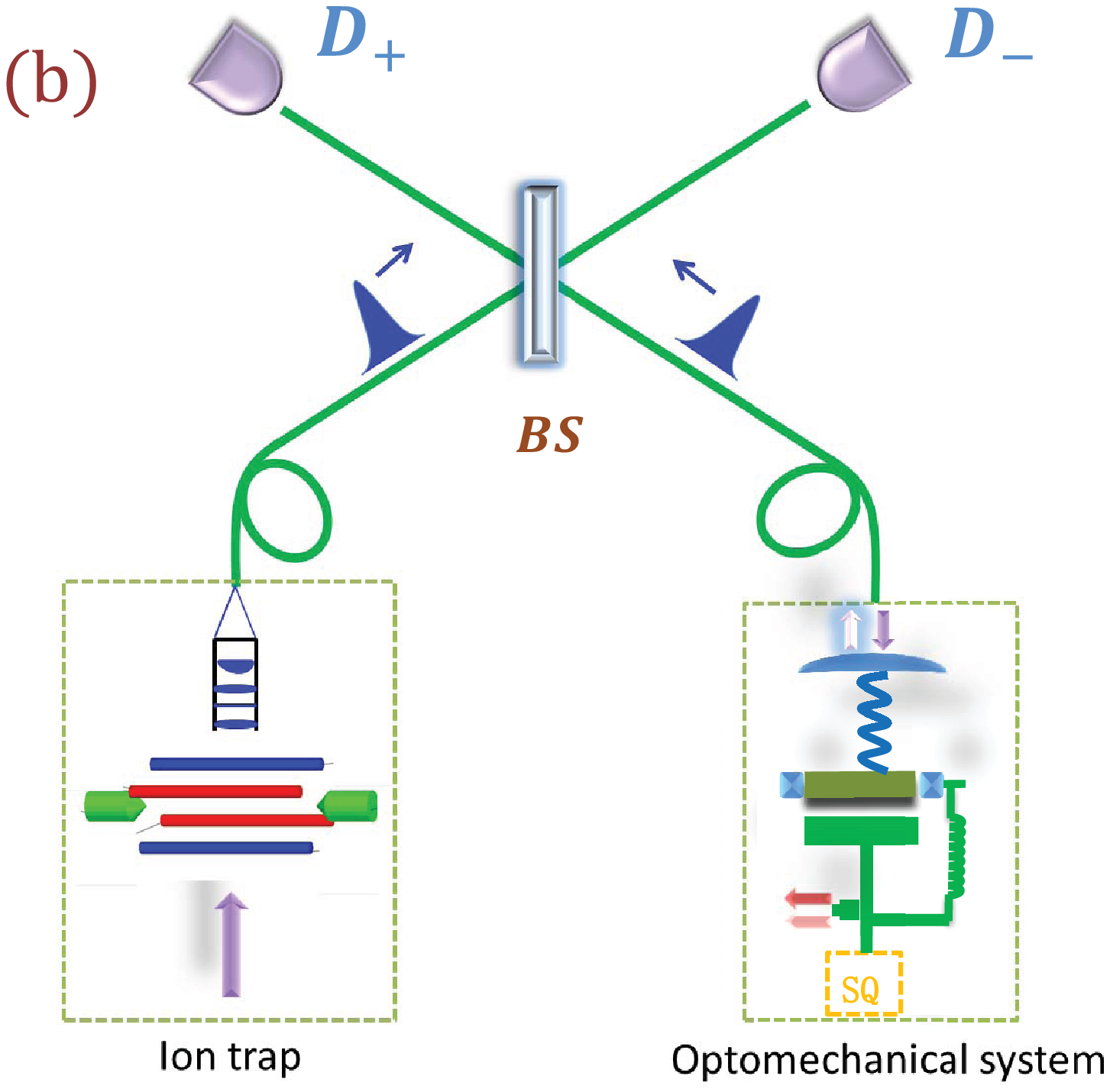}
\caption{(Color online) (a) The schematic to generate entanglement between 
remote SQs. Two SQs are located in distant cavities A and B. 
The SQs with dashed boxes represent the same structure as the orange part (SQ)
in Fig. 1a, capacitively coupled to the SRs. The SQs couple  
to the output photons through opto-mechanical interfaces. The output
photons, after propagation, interfere at a beam splitter and then
are detected by single-photon counters. Registration of a photon-count
generates entanglement between the remote SQs. (b) The same setup
can be used to entangle SQs with other kinds of matter qubits, such
as trapped ions. The carrier frequency and shape of the photon from
the SQ is tuned by the opto-mechanical interface to match with the
photon pulse from other matter qubits.}
\label{fig:entangle}
\end{figure}

The entanglement between remote SQs can also be generated in a probabilistic
fashion through detection of interference of the emitted photon(s)
\cite{Duan03,Barrett05,Galland14}. For instance, as shown in Fig. 4a, we have
SQs in two remote cavities, each emitting a single-photon pulse with
a small probability $p_{0}=1-\exp[-\kappa\int_{0}^{T_D}\sin^{2}\theta(\tau)d\tau]$
through an incomplete adiabatic passage from the state $|s\rangle$
to $|g\rangle$. The emitted pulses, after propagation in optical
channels, interfere at a $50-50\%$ beam splitter, with outputs detected
by single-photon counters. If we register only one photon from these
detectors, the two SQs are projected to an entangled state $\left(|s\rangle_{1}|g\rangle_{2}+e^{i\varphi}|g\rangle_{1}|s\rangle_{p}\right)/\sqrt{2}$
with a success probability proportional to $p_{0}\ll1$. The unknown
relative phase $\varphi$ can be canceled during the detection process
\cite{Chou05}, or through the second
round of entanglement generation by applying the same protocol again
\cite{DLCZ}. Compared with the deterministic scheme \cite{Cirac97},
this probabilistic scheme has a lower efficiency as the protocol needs
to be repeated until one successfully registers a photon count, however,
it is more robust to noise as the photon loss in the optical channels
does not influence the fidelity of this scheme.

A major challenge for quantum networking based on the photonic connection
is to achieve the spectrum (shape) and frequency matching of the emitted
photon pulses from different matter qubits. For solid-state qubits
in particular, the coupling parameters usually vary for different
systems and it is hard to get identical qubits or coupling rates.
A remarkable advantage of the scheme based on the opto-mechanical
interface is that all the mismatches in frequencies or pulse shapes
can be easily compensated through the driving fields. For instance,
the scheme works perfectly well if the coupling or decay rates are
different for different systems. As the pulse shape only depends on
$\theta(t)$ from Eq. (4), we can always get identical shapes as difference
in the coupling rates can be easily compensated by the microwave driving
amplitude $\Omega(t)$. Furthermore, the output optical frequency
is purely determined by the eigenmode structure of the optical cavity
and not limited by the qubit parameters. So, depending on the frequency
and shape of the driving field, we can have a quantum interface between
the SQ and the optical photon with widely tunable carrier frequency
and shape, which can then interfere with the photons emitted by other
kinds of matter qubits, such as trapped ions \cite{DuanRMP}, quantum dots
\cite{Schaibley13,Gao12}, or diamond nitrogen vacancy centers \cite{Bernien13}.
The SQ-opto-mechanical interface therefore can work as a quantum transducer
to generate entanglement links between different types of matter qubits.
This leads to a hybrid quantum network, with an example illustrated
in Fig. 3(b), which has the important advantage to combine the particular
strength of each kind of matter qubits.

\section{SQ initialization fidelity and interface efficiency}
In the above analysis, we assume the SQ couples dominantly
to the output field of the optical cavity and neglect other dissipation
channels. Now we take into account all the other dissipation processes
and calculate their effects on the fidelity of quantum interface.
Under the condition that the pulse duration $T_{D}^{-1}\ll G,g,\kappa_{1}$,
we can adiabatically eliminate all the modes $a_{j}$ ($j=1,2,m$)
in the Langevin equations (3) and arrive at the following decay equation
for the SQ (see details in Appendix \ref{appendixB}):
\begin{equation}
\dot{\sigma}_{ge}=-\frac{\gamma_{eff}}{2}\sigma_{ge}+\sqrt{\gamma_{eff}}\sigma_{z}a_{eff}^{\mathrm{in}},\label{eq:LangevinS}
\end{equation}
where $\gamma_{eff}=\gamma+\tilde{\kappa}_{1}+\tilde{\kappa}_{2}+\tilde{\kappa}_{m}$,
$a_{eff}^{in}=[-i\sqrt{\tilde{\kappa}_{1}}a_{1}^{in}+i\sqrt{\tilde{\kappa}_{2}}a_{2}^{in}+\sqrt{\gamma}a_{s}^{in}+\sqrt{\tilde{\kappa}_{m}}a_{m}^{in}]/\sqrt{\gamma_{eff}}$,
$\tilde{\kappa}_{1}=\frac{4g^{2}\kappa_{1}}{(\kappa_{1}+\kappa_{2}+\kappa_{1}\kappa_{2}\kappa_{m}/4G^{2})^{2}}$,
$\tilde{\kappa}_{2}=\frac{(2+\kappa_{1}\kappa_{m}/2G^{2})^{2}g^{2}\kappa_{2}}{(\kappa_{1}+\kappa_{2}+\kappa_{1}\kappa_{2}\kappa_{m}/4G^{2})^{2}}$,
and $\tilde{\kappa}_{m}=\frac{g^{2}\kappa_{1}^{2}\kappa_{m}/G^{2}}{(\kappa_{1}+\kappa_{2}+\kappa_{1}\kappa_{2}\kappa_{m}/4G^{2})^{2}}$.
The physical meaning of Eq. (5) is clear: the SQ couples to four decay
channels, the optical channel $a_{1}^{in}$ with decay rate $\tilde{\kappa}_{1}$,
the microwave channel $a_{2}^{in}$ with decay rate $\tilde{\kappa}_{2}$,
the mechanical channel $a_{m}^{in}$ with decay rate $\tilde{\kappa}_{m}$,
and the intrinsic channel $a_{s}^{in}$ with decay rate $\gamma$.
For each decay channel, the effective dissipation rate is given by
$\left(\bar{n}_{j}+1\right)\tilde{\kappa}_{j}$, where $\bar{n}_{j}=1/(\exp(\hbar\omega_{j}/k_{B}T)-1)$
is the mean thermal photon (or phonon) number and $T$ denotes temperature
of the system. The initialization of the
SQ is described by the Langevin equation
(5) and the final probability $P_{g}$ for the SQ in the state $|g\rangle$
is determined by the stationary state under Eq. \eqref{eq:LangevinS} (after a decay
time of the order of $1/\tilde{\kappa}_{1}\sim10\: ns$ with
\begin{equation}
P_{g}=\frac{\tilde{\kappa}_{1}+(n_{2}+1)\tilde{\kappa}_{2}+(n_{m}+1)\tilde{\kappa}_{m}}{\tilde{\kappa}_{1}+(2n_{2}+1)(\gamma +\tilde{\kappa}_{2})+(2n_{m}+1)\tilde{\kappa}_{m}}.
\end{equation}
Under experimental parameters list in the caption of Fig. (4) and
$1$ K system temperature, the fidelity $P_{g}$ for state initialization
is larger than $99\%$ (we assume temperature $T=1$ K with
$\bar{n}_{2}=1.62$, and $\bar{n}_{m}=2.08\times 10^3$).

\begin{figure}[htbp]
\centering \includegraphics[width=4.25cm]{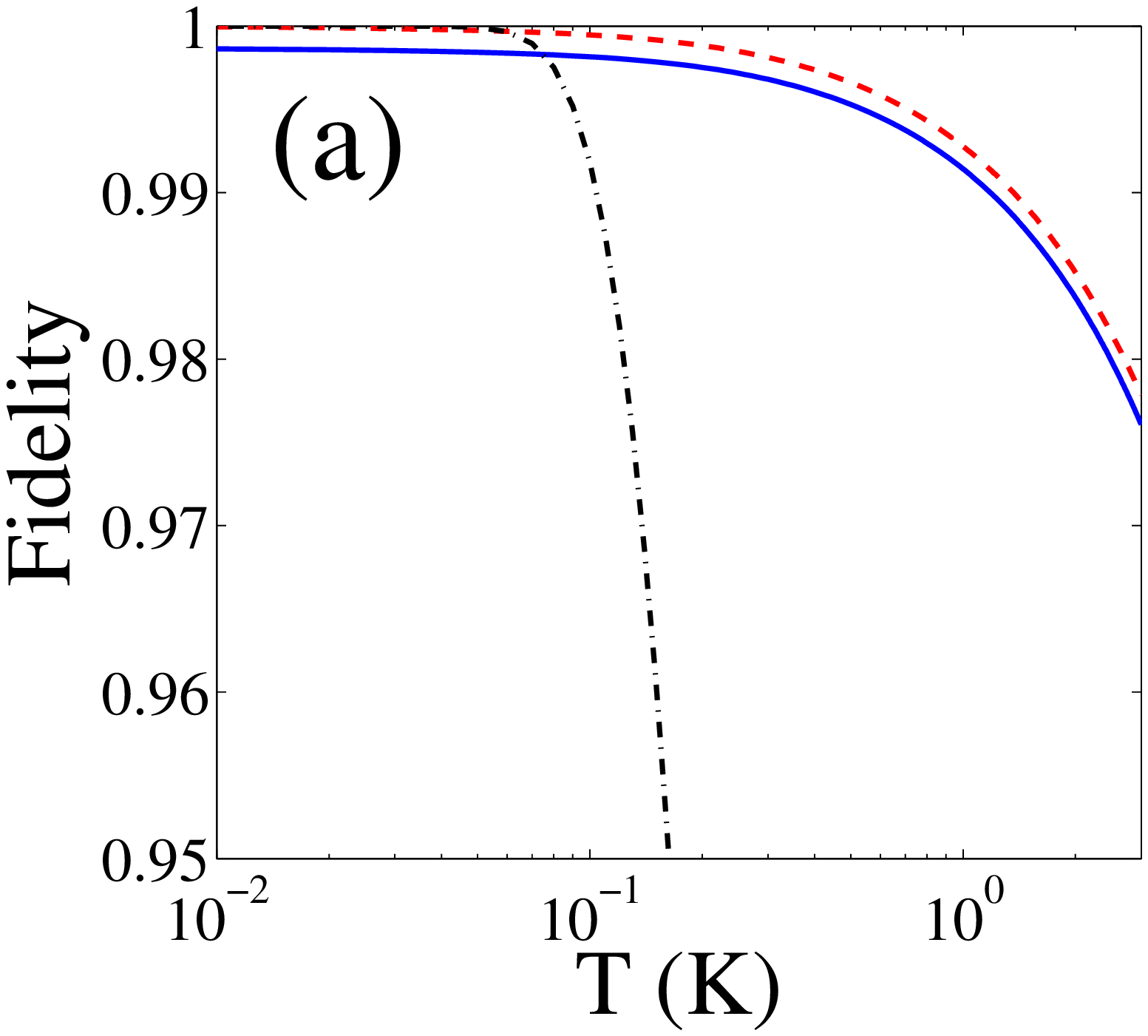} \includegraphics[width=4.25cm]{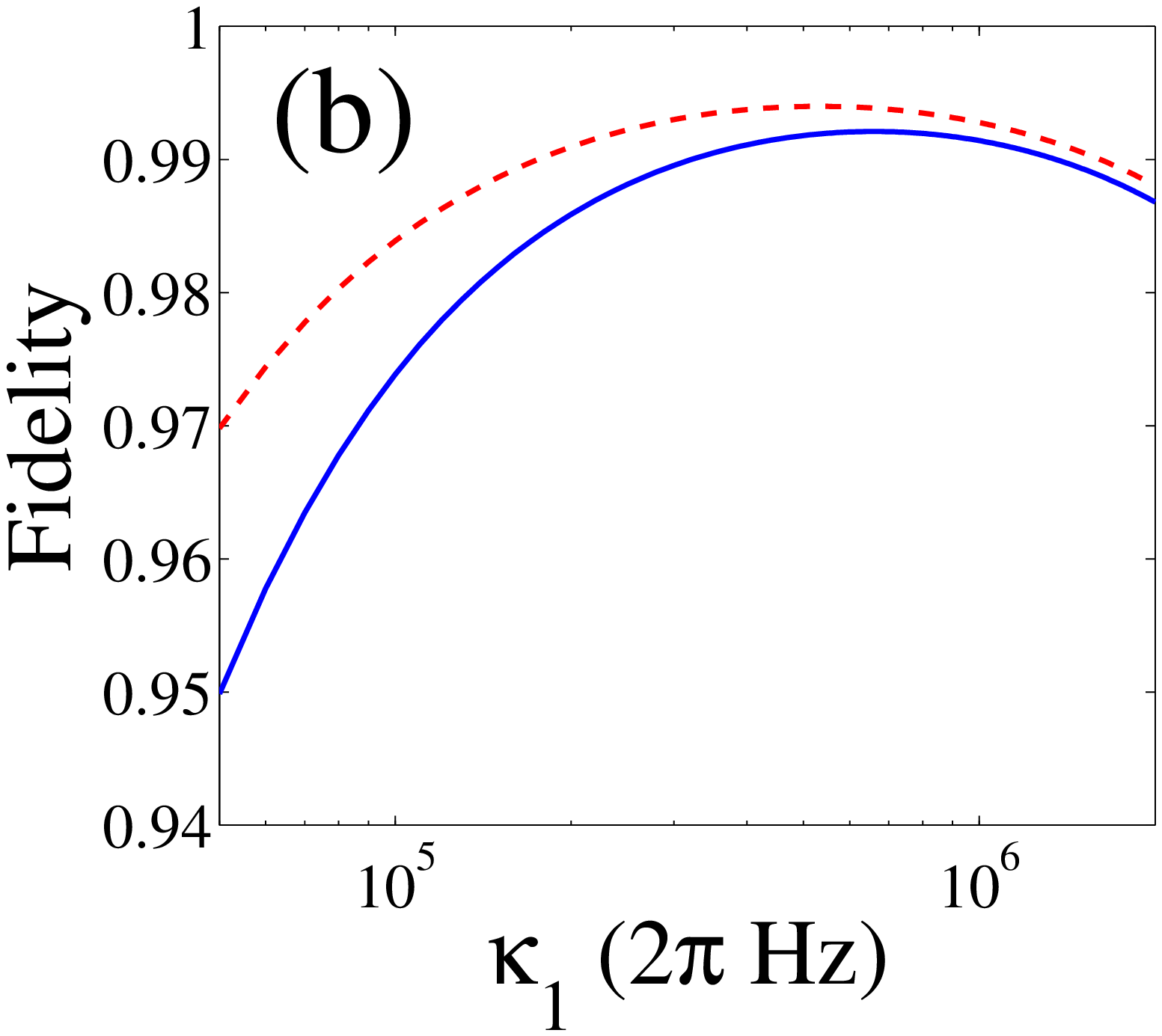}
\caption{(Color online) (a) The temperature dependence of the fidelity $F$ (solid curve)
of quantum interface and the fidelity $P_{g}$ (dashed curve) for
state initialization. The dash-dotted curve shows the probability in the
ground state without opto-mechanical sideband cooling. The parameters
are taken as $\omega_{1}/2\pi=200$ THz, $\omega_{2}/2\pi=10$ GHz,
and $\omega_{m}/2\pi=10$ MHz \cite{Teufel11a,Teufel11b,Pirk12},
$\kappa_{1}/2\pi=10$ MHz, $\kappa_{2}/2\pi=1$ kHz, $\kappa_{m}/2\pi=10$
Hz \cite{Palomaki13}, $\gamma/2\pi=5$ kHz \cite{Rigetti12,Reagor13}, $G/2\pi=1$
MHz, and $g_c/2\pi=1$ MHz. (b) The dependence of the fidelity $F$
(solid curve) and $P_{g}$ (dashed curve) on the optical cavity decay
rate $\kappa_{1}$ at $1$ K temperature. The other parameters are
the same as Fig. (a).}

\label{fig:fidelity}
\end{figure}

%

For quantum networking of SQs through the optical decay
channel, all the other dissipation channels contribute to noise, and
the fidelity $F$ of the quantum interface can be estimated by the
relative ratio of the optical decay rate to the total dissipation
rate
\begin{equation}
F=\frac{\tilde{\kappa}_{1}}{\tilde{\kappa}_{1}+(\bar{n}_{2}+1)\tilde{\kappa}_{2}+(\bar{n}_{m}+1)\tilde{\kappa}_{m}+(\bar{n}_{2}+1)\gamma},
\end{equation}
where we have taken $\bar{n}_{1}\thickapprox0$ at the optical frequency.
The experimental parameters typically satisfy $G\sim\kappa_{1}\gg\kappa_{2},\kappa_{m},\gamma$.
In this case, $\tilde{\kappa}_{1}\approx4g^{2}/\kappa_{1}$, $\tilde{\kappa}_{2}\approx 4g^{2}\kappa_{2}/\kappa_{1}^{2}$,
and $\tilde{\kappa}_{m}\approx g^{2}\kappa_{m}/G^{2}$. In Fig. 4,
we show the fidelity as a function of the system temperature and the
decay rate of the optical cavity. It is found that the fidelity is
around $99\%$ under typical values of the experimental parameters
as listed in the figure caption.

Typically the SQ system is operated around $20$ mK temperature, where
the ground state cooling is achieved directly through thermal equilibrium.
However, with an opto-mechanical interface, the system temperature
may increase due to heating by the black-body radiation from the optical
window. Here, we show that even under $1$ K temperature, the state
can still be initialized through the opto-mechanical sideband cooling.
Another requirement for the system temperature is that the quasi-particle
density in the superconducting circuit should be small, otherwise
it will induce dissipation of the SQ. The quasi-particle density is
proportional to $e^{-1.76T_{c}/T}$, where $T_{c}$ is the critical
temperature of the superconductor \cite{BCS57}. For niobium, the
critical temperature $T_{c}$ is about $9.3$ K, for which the quasi-particle
density is negligible at $1$ K temperature. For aluminum, the $T_{c}$
is about $1.2$ K, where the quasi-particles can be neglected only
at temperature in the order of $0.1$ K.

In summary, we have proposed a scheme to realize a quantum network
of SQs base on the opto-mechanical quantum interface. The interface
can couple the SQs to optical photons with widely tunable carrier
frequencies and pulse shapes. The same interface can also be used
for fast initialization of the SQs at $1$ K temperature through opto-mechanical sideband
cooling.

This work was funded by the NBRPC (973 Program) 2011CBA00300 (2011CBA00302),
NNSFC 11105136, 11474177, 61435007. WLY. was is supported by the
National Fundamental Research Program of China under Grant No. 2012CB922102
and by the NNSFC 11274351. LMD acknowledge support from the IARPA
MUSIQC program, the ARO and the AFOSR MURI program.

\appendix
\section{Effective linear Hamiltonian} \label{appendixA}
The Hamiltonian of the system takes the form $H=H_{0}+H_{I}+H_{drive}$,
where
$$H_{0}=\sum_{i=1,2}\omega_{i}a_{i}^{\dagger}a_{i}+\omega_{m}a_{m}^{\dagger}a_{m}+\omega_{e}\sigma_{ee},$$
$$H_{I}=\sum_{i=1,2}g_{i}a_{i}^{\dagger}a_{i}(a_{m}+a_{m}^{\dagger})+g_{c}(\sigma_{eg}+\sigma_{ge})(a_{2}+a_2^\dagger),$$
and $$H_{d}=\sum_{i=1,2}(\frac{\Omega_{i}}{2}e^{-i\omega_{Li}t}+ \mathrm{h.c.})(a_i+a_i^\dagger)+ (\frac{\Omega'}{2}
e^{-i\omega_{L_2} t} + \mathrm{h.c.})( \sigma_{ge}+\sigma_{eg}).$$
The SQ is assumed to couple resonantly with the SR with $\omega_{2}=\omega_{e}$.
The detuning $\Delta_i=\omega_i-\omega_{L_i}=\omega_m$.
Under the condition that $\Omega_i < 4\omega_i$, the Hamiltonian $H_d$ can be approximated as
$$H_d'=\sum_{i=1,2}(\frac{\Omega_{i}}{2} a_i e^{-i\omega_{Li}t}+ \mathrm{h.c.})+ (\frac{\Omega'}{2}
e^{-i\omega_{L_2} t} + \mathrm{h.c.})( \sigma_{ge}+\sigma_{eg}).$$
We take the rotating wave frame that $H_0'=H_0-\omega_m (a_1^\dagger a_1+
a_2^\dagger a_2$. The Hamiltonian in rotating wave frame reads
\begin{equation}\label{eq:Hr}
\begin{aligned}
  H_R=& \omega_m \sum_{p=1,2,m} a_p^\dagger a_p +\sum_{i=1,2}[ g_i a_i^\dagger a_i (a_m+a_m^\dagger)
  +(\frac{\Omega_i}{2} a_i + \mathrm{h.c.})] \\&+\omega_e \sigma_{ee} + (g_c  a_2 e^{-i\omega_{L_2}t} + \frac{\Omega'}{2} e^{-i\omega_{L_2}t} + \mathrm{h.c.})
  (\sigma_{ge}+\sigma_{eg})
  \end{aligned}
\end{equation}

We assume that the decay rates $\kappa_i$ for mode $a_i$ (i=1,2) are much less than the driving detuning $\Delta=\omega_m$.
Under the driving, the steady state amplitude of the mode $a_{i}$ is given by $\alpha_{i}\simeq \Omega_{i}/2\omega_m$.
In order to compensate the effect of classical driving on SQ, we set $\Omega'^*=2\alpha_2 g_c =\Omega_2 g_c/\omega_m$.
In the limit that $\alpha_i \gg 1$, the Hamiltonian Eq. \eqref{eq:Hr} can be expanded with $a_{i}-\alpha_{i}$
\begin{equation}
 \begin{aligned}
H_{om}=& \sum_{i=1,2}\left[\omega_m a_{i}^{\dagger}a_{i}+G_{i}(a_{i}^{\dagger}+a_{i})(a_{m}+ a_{m}^{\dagger})\right] + \omega_{e} \sigma_{ee} \\&+\omega_{m}a_{m}^{\dagger}a_{m}+( g_c a_2 e^{-\omega_{L_2}t}+\mathrm{h.c.})(\sigma_{eg}+\sigma_{ge}).
 \end{aligned}
\end{equation}
 Under the rotating wave approximation ($\omega_{m}\gg G_{i}, g_c$),
the whole Hamiltonian in the interaction picture is given by
\begin{equation}
H_{I}=\left(G_{1}a_{1}^{\dagger}+G_{2}a_{2}^{\dagger}\right)a_{m}+g_{c}\sigma_{eg}a_{2}+\mathrm{h.c.}.
\end{equation}

Here we take the parameters  we used in Fig. 4 as an example
to make sure that the rotating  wave approximation is valid.
In experiments, the typical parameters are as follows: $\omega_{1}/2\pi=200$ THz, $\omega_{2}/2\pi=10$ GHz,
and $\omega_{m}/2\pi=10$ MHz \cite{Teufel11a,Teufel11b,Pirk12},
$\kappa_{1}/2\pi=10$ MHz, $\kappa_{2}/2\pi=1$ kHz, $\kappa_{m}/2\pi=10$
Hz, $\gamma/2\pi=5$ kHz \cite{Rigetti12,Reagor13}, $g/2\pi=1$
kHz, and $g_c/2\pi=1$ MHz. The microwave driving strengths are assumed to be $\Omega=20$ GHz. The steady state
amplitude $\alpha=1000$ and $\Omega'=\Omega*g_c/\omega_m=2$ GHz. The effective coupling between $a_2$ and
$a_m$ is $G_2= \alpha_2 g_2= 1$ MHz. With proper driving on optical
cavity mode $a_1$, we can also get the effective coupling strength $G_1=1$ MHz.
Therefore rotating wave approximation condition
$\omega_m \gg G_i, g_2$ is fulfilled.

\section{Effective Langevin Equation for SQ.} \label{appendixB}

In order to derive the effective Langevin Equation for the SQ, we write down the Langevin equations of the systems
\begin{eqnarray}
  \dot{a_{1}} & = & -i G a_{m } - \frac{\kappa_{1}}{2} a_{1} +
  \sqrt{\kappa_{1}} a_{1}^{\tmop{in}}  \label{a1}\\
  \dot{a}_{2} & = & -i G a_{m} +i g \sigma_{\tmop{ge}} - \frac{\kappa_{2}}{2}
  a_{2} + \sqrt{k_{2}} a_{2}^{\tmop{in}}  \label{a2}\\
  \dot{a}_{m} & = & -i G  ( a_{1} +a_{2} ) - \frac{\gamma_{m}}{2} a_{m} +
  \sqrt{\kappa_{m}} a_{m}^{\tmop{in}}  \label{am}\\
  \dot{\sigma}_{g e} & = & i g \sigma_{z} a_{2}  - \frac{\gamma}{2} \sigma_{g
  e} + \sqrt{\gamma} \sigma_{z} \sigma^{\tmop{in}}_{g e}  \label{s}
\end{eqnarray}
In the limit that $G \gg g, \kappa_{1} , \kappa_{2} , \kappa_{m}$, we can
adiabatically eliminate $a_{m}$ and $a_{1,2}$ modes. Let's solve $a_{m}$ from
Eq. (\ref{a1}) in term of $a_{1}$,
\begin{equation}
a_{m} = \frac{1}{i G} \left( - \frac{\kappa_{1}}{2}  a_{1} +
   \sqrt{\kappa_{1}} a_{1}^{\tmop{in}} \right) , \label{am1}
\end{equation}
Then we can solve the $a_{2}$ from Eq. (\ref{a2}) in term of $a_{1}$
\begin{eqnarray*}
  a_{2}    =  \frac{\kappa_{1}}{\kappa_{2}} a_{1} - \frac{2}{\kappa_{2}}
  \sqrt{\kappa_{1}} a_{1}^{\tmop{in}} + \frac{2i g \sigma_{g e}}{\kappa_{2}} +
  \frac{2}{\sqrt{\kappa_{2}}} a_{2}^{\tmop{in}}.
\end{eqnarray*}
Let's solve $a_{1}$ from Eq. (\ref{am}) and get the expression of $a_{2}$,
\begin{eqnarray*}
  a_{1}  =  \frac{1}{i \left(  G+ \frac{\gamma_{m} \kappa_{1}}{4G} \right)}
  \left( -i G a_{2} + \frac{i \gamma_{m}}{2G} \sqrt{\kappa_{1}}
  a_{1}^{\tmop{in}} + \sqrt{\gamma_{m}} a_{m}^{\tmop{in}} \right).
\end{eqnarray*}
Inserting $a_{1}$ into the expression of $a_{2}$, we get that
\begin{equation}
 \begin{aligned}
  a_{2}  =&   \frac{-8G^{2} \sqrt{\kappa_{1}}}{\kappa_{2} ( 4G^{2} + \kappa_{m}
  \kappa_{1} )} a_{1}^{\tmop{in}} - \frac{4i G \kappa_{1}
  \sqrt{\kappa_{m}}}{\kappa_{2} ( 4G^{2} + \kappa_{m} \kappa_{1} )}
  a_{m}^{\tmop{in}} \\&+ \frac{2i g \sigma_{g e}}{\kappa_{2}} +
  \frac{2}{\sqrt{\kappa_{2}}} a_{2}^{\tmop{in}}
  \end{aligned}
\end{equation}
We get that
\begin{equation}
 \begin{aligned}
  a_{2}   =& \frac{1}{4G^{2} ( \kappa_{1} + \kappa_{2} ) + \gamma_{m} \kappa_{1}
  \kappa_{2}} [ -8G^{2} \sqrt{\kappa_{1}} a_{1}^{\tmop{in}} - 4i G \kappa_{1}
  \sqrt{\kappa_{m}} a_{m}^{\tmop{in}} \\&+ ( 8G^{2} +2
  \kappa_{m} \kappa_{1} ) \sqrt{\kappa_{2}} a_{2}^{\tmop{in}}  +i ( 8G^{2} +2 \kappa_{m} \kappa_{1} ) g
  \sigma_{g e} ]
  \end{aligned}
\end{equation}
Finally we get the effective Langevin equation for $\sigma_{g e}$ is
\begin{eqnarray}
  \dot{\sigma}_{g e}  &=&  - \left( \frac{2g^{2} +g^{2} \kappa_{m} \kappa_{1} /2G^{2}}{(
  \kappa_{1} + \kappa_{2} ) + \kappa_{m} \kappa_{1} \kappa_{2} /4G^{2}} +
  \frac{\gamma}{2} \right) \sigma_{g e} \nonumber \\&& + \frac{-2i g \sqrt{\kappa_{1}}
  \sigma_{z}}{( \kappa_{1} + \kappa_{2} ) + \kappa_{m} \kappa_{1} \kappa_{2}
  /4G^{2}} a_{1}^{\tmop{in}} \nonumber \\&& + \frac{i ( 2+ \kappa_{m} \kappa_{1} /2G^{2} ) g
  \sqrt{\kappa_{2}} \sigma_{z}}{( \kappa_{1} + \kappa_{2} ) + \kappa_{m}
  \kappa_{1} \kappa_{2} /4G^{2}} a_{2}^{\tmop{in}} \nonumber \\&& + \frac{g \kappa_{1}
  \sqrt{\kappa_{m}} /G}{( \kappa_{1} + \kappa_{2} ) + \kappa_{m} \kappa_{1}
  \kappa_{2} /4G^{2}} \sigma_{z} a_{m}^{\tmop{in}} + \sqrt{\gamma} \sigma_{z}
  \sigma_{g e}^{\tmop{in}}  \label{seff}
\end{eqnarray}
 It is easy to verify that the effective Langevin equation \eqref{seff} fulfills  the Einstein relation.

The effective Langevin equation \eqref{seff} for $\sigma_{g e}$ can be rewritten as
\begin{eqnarray}
\dot{\sigma}_{g e} =- \frac{\gamma_{\tmop{eff}}}{2} \sigma_{g e} +
   \sqrt{\gamma_{\tmop{eff}}} \sigma_{z} a_{\tmop{eff}}^{\tmop{in}} ,
   \end{eqnarray}
where
$\gamma_{\tmop{eff}}$ and $a_{\tmop{eff}}^{\tmop{in}}$ are defined as
$$ \gamma_{\tmop{eff}}  =  \gamma + \frac{16G^{2} g^{2} +4g^{2} \kappa_{m}
  \kappa_{1}^{2}}{4G^{2} ( \kappa_{1} + \kappa_{2} ) + \kappa_{m} \kappa_{1}
  \kappa_{2}},$$ and
\begin{equation}
  \begin{aligned}
  a_{\tmop{eff}}^{\tmop{in}}  = & [ \frac{-2i g \sqrt{\kappa_{1}}}{(
  \kappa_{1} + \kappa_{2} ) + \kappa_{m} \kappa_{1} \kappa_{2} /4G^{2}}
  a_{1}^{\tmop{in}} + \frac{i ( 2+ \kappa_{m} \kappa_{1} /2G^{2} ) g
  \sqrt{\kappa_{2}}}{( \kappa_{1} + \kappa_{2} ) + \kappa_{m} \kappa_{1}
  \kappa_{2} /4G^{2}} a_{2}^{\tmop{in}} \\&+  \frac{g \kappa_{1} \sqrt{\kappa_{m}}
  /G}{( \kappa_{1} + \kappa_{2} ) + \kappa_{m} \kappa_{1} \kappa_{2} /4G^{2}}
  a_{m}^{\tmop{in}} + \sqrt{\gamma} a_{2}^{\tmop{in}} ] /
  \sqrt{\gamma_{\tmop{eff}}}.
  \end{aligned}
\end{equation}

\end{document}